\documentclass[12pt]{iopart}
\usepackage[pdftex]{graphicx}
\usepackage{dcolumn}% Align table columns on decimal point
\usepackage{bm}% bold math
\usepackage{array}
\usepackage{color}
\usepackage{float}
\usepackage{subfigure}
\usepackage{dsfont}
\usepackage{txfonts}
\usepackage{wasysym}
\usepackage{multirow}
\usepackage{sidecap}
\usepackage{xcolor,cancel}
\usepackage[normalem]{ulem}
\usepackage{cite}
\usepackage{lipsum}
\usepackage{cuted}
\usepackage{flushend}

\newcommand{\+}{\dagger}

\newcommand{\dn}{\downarrow}

\newcommand{\wire}{\mathrm{wire}}

\newcommand{\up}{\uparrow}
\newcommand{\down}{\downarrow}

\newcommand{\RKKY}{{\rm RKKY}}
\newcommand{\imp}{{\rm imp}}
\newcommand{\Is}{{\rm Ising}}

\newcommand{\DM}{{\rm DM}}
\renewcommand{\Re}{{\rm Re \,}}
\renewcommand{\Im}{{\rm Im \,}}

\begin{document}

\title{Modified exchange interaction between magnetic impurities in spin-orbit 
coupled quantum wires}

\author{Joelson F. Silva and E. Vernek}

\address{Instituto de F\'isica, Universidade Federal de Uberl\^andia, 
Uberl\^andia, Minas Gerais 38400-902, Brazil.}
\ead{vernek@ufu.br}
\date{\today}
\begin{abstract}
Indirect exchange interaction between magnetic impurities in one dimensional 
systems is a matter of long discussions since Kittel has established that in  
the asymptotic limit  it decays as the inverse of distance $x$ between the 
impurities. In this work we investigate  this problem in a quantum wire with 
Rashba spin-orbit coupling (SOC). By employing a second order perturbation 
theory we find that one additional oscillatory term appears 
in  each of the RKKY, the Dzaloshinkii-Moryia and the Ising couplings. 
Remarkably, these extra terms resulting from the spin  precession of the 
conduction electrons induced by the SOC  do not decay as in the usual RKKY 
interaction.  We show that these extra oscillations arise from  the finite 
momenta band splitting  induced by the spin-orbit coupling that  modifies the  
spin-flip scatterings occurring at the Fermi energy. Our findings open up a new 
perspective in the long-distance magnetic interactions in solid state 
systems.
\end{abstract}

\maketitle
%\ioptwocol
% +|\gamma|(k-k^\prime
% 

\section{Introduction}
Indirect exchange interactions among magnetic impurities embedded in conduction 
electrons is  a rich and fascinating problem in solid state physics. The most 
familiar inter-impurity interaction  mediated by the conduction electrons 
is the celebrated Ruderman-Kittel-Kasuya-Yosida (RKKY) 
interaction~\cite{Kittel_book}.  The discovery of the the RKKY interaction 
allowed for the comprehension of magnetic order of a variety magnetic 
materials~\cite{Vleck}. This phenomena can be  understood within the 
concept of perturbation theory: an electron scattered by a given magnetic 
impurity has its spin modified by a local exchange interaction ---a Kondo-like 
coupling. This information is then transfered to a second impurity upon a 
second similar collision. The net effect is an effective indirect 
exchange coupling between the two impurities mediated by the conduction 
electrons~\cite{Ruderman,Kasuya,Yosida}. In conventional systems, this 
resulting  effective coupling exhibits an oscillatory  behavior as 
a function of the distance $x$ between the impurities, decaying as $x^{-D}$, 
where $D$ is the dimension of the system. 

In the recent years we have witnessed a renewed interest in the indirect 
exchange interactions between magnetic impurities embedded in spin-orbit 
coupled conduction electrons~\cite{Imamura,Egger,Kandu,Wang}, including 
topological insulators, Dirac~\cite{Mastrogiuseppe} and Weyl~\cite{Hao-Ran} 
semimetals. In these materials, the spins of the conduction electrons and their 
momenta are coupled together. As a result,  after been scattered by 
the one impurity, the  spin of a given electron precesses while traveling 
towards the second impurity. This precession produces a more complex and 
reacher inter-impurity magnetic interaction\cite{Qing,Loss} such as twisted magnetic 
arrangement, a non-collinear exchange coupling known as Dzaloshinkii-Moryia 
interaction (DMI)~\cite{Dzyaloshinsky,Moriya} and a Ising like 
coupling~\cite{Mross}.  From a practical point of view, the Rashba spin-orbit 
coupling (SOC) opens up the possibility of controlling the inter-impurity 
magnetic interaction via external electric field with great potential 
application in spintronics~\cite{Spintronics,Tian}. Particularly appealing, but 
hitherto less investigated, is the indirect the exchange interactions in 1D 
systems in the presence SOC. Since in 1D the electrons are forced to propagate 
along some particular direction, the spin-momentum locking induced by the SOC 
can drastically modify the scattering processes~\cite{Sousa}. 
 
\begin{figure}[h!]
\centering
\subfigure{\includegraphics[clip,width=4.5in]{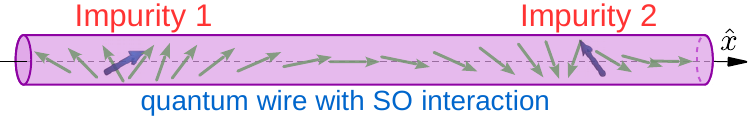}}
\caption{Schematic representation of the system composed of two magnetic 
impurities coupled to a quantum wire with spin-orbit interaction. Blue
arrows represent the magnetic moments of the impurities while green arrows 
represent the spins of the conduction electrons that precess due to the 
spin-orbit coupling.} 
\label{model}
\end{figure} 

In a seminal paper published in 1990, Datta and Das  proposed the idea 
of producing a highly spin-polarized current controlled via SOC by external electric field~\cite{Datta_Das}. In their device, the spins of the electrons injected from a polarized source could be rotated by a tunable SOC while traveling  towards a polarized drain. Likewise, it would interesting if one 
could use the SOC to control the indirect exchange interaction between two  
magnetic impurities embedded in a 1D conduction electron sea. 
The few studies  addressing the  RKKY interaction in one-dimensional 
systems with SOC, in general,  employ a real space Green's 
function~\cite{Imamura}. It is known, however, that calculating the RKKY 
interaction in one-dimension is quite subtle~\cite{Giuliani}. This was 
first noticed by Kittel~\cite{Kittel_book_1} and latter discussed in detail by 
Yafet~\cite{Yafet}.  Yafet indeed showed that, depending on how the double 
integral is handled,  it can  lead to  unphysical results. Moreover, Yafet 
observed that the problem arises because the Pauli's exclusion principle is
severely violated. More recently, Rusin and Zawadzki~\cite{Zawadzki} has 
examined the commonly used expression for the RKKY interaction~\cite{Imamura} 
and noticed that there is an implicit change  of order in a double-integration 
that requires extra care when used in one dimensional cases. Motivated  by the 
interest in the physics of the RKKY interaction renewed in SOC materials,  we 
revisit the calculation of the full indirect exchange interaction between two  
magnetic impurities in a 1D system. Based on the traditional second order 
perturbation theory, we obtain the known form of the inter-impurity couplings, 
which includes the usual RKKY, DM and Ising interaction terms. The 
effective couplings are calculated both numerically and analytically. 
Drastically different from  the usual RKKY systems,  we obtain additional  
oscillatory contributions to the effective couplings  that do not decay with the 
distance.  These unsuppressed terms vanish in the absence of SOC, in which case 
the traditional RKKY coupling is recovered.  This feature is 
potentially important to spintronics as it can be used to control spin-spin 
interaction at longer distances as compared to the traditional RKKY couplings.  
Indeed, by employing a similar calculation we perform here, it was shown important enhancement in the magnetic coupling between magnetic impurity in controlled Rashba spin-orbit interaction\cite{Lyua}. 
There are currently several modern 1D systems with SOC that are natural 
candidates for experimental investigation of this  interesting 
physics~\cite{Gao,Picciott,Spilman,Hirayama,Zwierlein}.

\section{Model and method}
We consider two spin-$1/2$ magnetic impurities\cite{footenote1} coupled to 
a quantum wire with spin-orbit interaction, as schematically shown in 
Fig.~\ref{model}. We write the full Hamiltonian of the  system  
as 
$H\!=\!H_{\wire}+H_{1}+H_2$, where 
\begin{equation}\label{H_RD_1}
H_{\wire}= \sum_k \left(\varepsilon_k\delta_{ss^\prime}
-\alpha_{\rm R} k\sigma^{y}_{s s^\prime}\right) c^\dagger_ { k s }c_{k 
s^\prime}
\end{equation}
describes the quantum wire, in which $c_{ks}^\+$ ($c_{ks}$) creates (annihilates) and electron with wave vector $k$,  spin $s$ and energy $\varepsilon_k$. Here, 
$\varepsilon_k=\hbar^2k^2/2m^*$, where $m^*$ is the effective mass of the 
conduction electrons. The linear Rashba~\cite{Rashba}  spin-orbit coupling is 
described by the term proportional to $\alpha_{\rm R}$, 
with $\sigma^\mu$ representing the $\mu$-th the Pauli matrix. Finally, the 
couplings between the impurities and the conduction band are given 
by~\cite{Nolting} 
\begin{eqnarray}\label{H_Kondo}
H_i&=&\frac{J}{N}\sum_{kk^\prime}e^{-i(k-k^\prime)
x_i}\left[S^z_i\left(c^\+_ { k^\prime\up } c_ { k\up } -c^\+_ { 
k^\prime\dn}c_{k\dn}\right)+ S^+_ic^\+_{k^\prime\dn}c_{k\up} 
+ S^-_ic^+_{k^\prime\up}c_{k\dn}\right],
\end{eqnarray}
where $x_i$ (with $i=1,2$) is the position of the $i-$th impurity.
We now derive an effective  coupling between the two 
impurities mediated by the conduction electrons. Starting by diagonalizing the 
Hamiltonian (\ref{H_RD_1}), we follow the traditional second 
order perturbation theory approach. The resulting inter-impurity interaction is 
described by the effective Hamiltonian~(see detail in \ref{App:Perturbation})
\begin{eqnarray}\label{eq10}
\tilde H_{\imp}&= &I_\parallel \tilde S_{1}^{z}\tilde S_{2}^{z} + 
\left(I_{\perp}\tilde S_{1}^{+}\tilde S_{2}^{-}+I_{\perp}^*\tilde 
S_{2}^{+}\tilde S_{1}^{-}\right).
\end{eqnarray}
The tildes on top of the spin operators above indicate that these 
operators are also written in the rotated basis. In Eq.(\ref{eq10}) 
we have defined {$I_{\parallel}\!=\!2\Re (I_{++}+I_{--})$ and 
$I_{\perp}\!=\!I_{-+}+I_{+-}^*$}, where
\begin{eqnarray}\label{coupling_integral0}
I_{\delta\nu}=\frac{J^2}{4\pi^2}\!\int_{-k_\delta}^{k_\delta}dk\int_{|k^\prime|>k_\delta } dk^\prime \frac{e^{i(k-k^\prime)x}  {
}} {\frac{\hbar^{2}}{2m^*}(k^{2} - 
k^{^\prime 2})+\delta\alpha_{\rm R}(k-\delta\nu k^\prime)}.
\end{eqnarray}
Here, $x=x_2-x_1$ is the distance between 
the impurities and $k_\delta=k_{\rm F}+\delta k_{\rm R}$, with 
$k_{\rm R}=m^*\alpha_{\rm R}/\hbar^2$ being the characteristic inverse of 
spin-orbit length. In the Eq.(\ref{coupling_integral0}) we also have  $\delta,\nu \in \{+,-\}$ denoting the Rashba bands. The rather simple form of the Hamiltonian (\ref{eq10}), written in the Rashba basis,  hides very interesting physics. It 
can be seen  that $I_\parallel \neq I_\perp $, therefore, in the present form, 
the Hamiltonian  (\ref{eq10}) describes a highly anisotropic exchange 
interaction mediated by the conduction electrons. To highlight  the physics 
buried in the Eq.~(\ref{eq10}) we transform it back to the original real 
spin basis,  obtaining 
\begin{eqnarray}\label{H_RKKY}
\tilde H_{\imp}&=&I_{\RKKY}\mathbf{S}_{1}\cdot 
\mathbf{S}_{2}+I_{\DM}[(\mathbf{S}_{1}\times \mathbf{S}_{2})\cdot 
\hat{\bf y}]+I_{\Is}S_{1}^{y}S_{2}^{y}.
\end{eqnarray}
Here, $I_{\RKKY}\!\!=2\Re I_{\perp}$, is the traditional RKKY interaction 
coupling renormalized by the SOC,  $I_{\DM}\!\!\!=\!\!\!-2\Im I_{\perp}$ is the 
Dzaloshinkii-Moryia interaction between the two impurities  and 
$I_{\Is}\!\!=\!\! I_{\parallel}-2\Re I_{\perp}$ represents an Ising-like 
coupling. Again, for $\alpha_{\rm R}\!\!=\!\!0$ only the 
first term of (\ref{H_RKKY}) survives. In this case we left with 
one double integral, obtaining $I_{\Is}\!=\!I_\DM\!=\!0$ and
\begin{eqnarray}\label{di}
I_{\RKKY}= 4I_0\int_{-k_{\rm F}}^{k_{\rm F}}{dk}\int_{|k^\prime|>k_{\rm F}}{
dk^\prime}\frac{\cos\left[(k-k^\prime)x\right] }{k^{2} - 
k^{\prime 2}},
\end{eqnarray}
with $I_0={m^* J^2 }/{2\pi^2\hbar^{2}}$.
Performing the double integral (\ref{di}) is known to be a delicate matter and 
have been discussed from way back \cite{Yafet}. Analytically, the integration 
can be performed if one extends the integral over $k^\prime$ to the entire real 
axis. After this, the residue theorem can be employed. Apart from the 
singular point  $k^\prime\!\!=\!\!k \!\!=\!\! 0$ (which can be accounted 
separately) the contribution to the double integral added by including the interval $(-k_{\rm F},k_{\rm F})$ vanishes  because the integrand is antisymmetric under exchange $k^\prime\leftrightarrow k$. In the asymptotic limit $x\rightarrow \infty$, the final correct solution exhibits the usual form $\cos(2k_{\rm F}x)/x$.  

In the presence of SOI $\alpha_{\rm R}\neq 0$, exact solutions 
for the integrals of Eq.~(\ref{coupling_integral0}) are, unfortunately, 
unavailable. In this case, even though we can subtract the 
contribution of the singularity from the integration over $k^\prime$ 
within the entire real axis, the integrand is no longer antisymmetric. 
Therefore, by extending the integral over $k^\prime$ to the interval 
$(-k_{\rm F},k_{\rm F})$, the extra contribution  cannot be fully subtracted. As we will see below, great approximate solutions  for the integrals 
(\ref{coupling_integral0}) can still be obtained in the limit $k_{\rm R} \ll 
k_{\rm F}$, in which case the asymmetry of the integrand is negligible.  To 
carry out the calculations, we simplify the notation  defining $a=2k_{\rm 
R}$ and $\tilde a =a/k_{\rm F}$ and introducing the dimensionless momenta  
$q=k/k_{\rm F}$ and $q^\prime=k^\prime/k_{\rm F}$. Within  these new 
variables,  we can rewrite the Eq.(\ref{coupling_integral0}) as
\begin{equation}\label{coupling_integral_1}
I_{\delta\nu}=I_0\int_{-q_\delta}^{q_\delta}{dq}
\int_ {|q^\prime|>q_\delta }\!\frac{e^{i(q-q^\prime)k_{\rm F}x} dq^\prime } 
{(q^{2} - q^{\prime 2})+\delta \tilde a(q-\delta\nu q^\prime)}.\nonumber \\
\end{equation}
Here, $q_\delta=1+\delta \tilde 
a/2=1+\delta k_{\rm R}/k_{\rm F}$. Following Yafet's 
approach~\cite{Yafet} we can write 
$I_{\delta\nu}=I_{\delta\nu}^r - I_{\delta\nu}^\epsilon$, where
\begin{eqnarray}\label{Ir0}
I_{\delta\nu}^r=I_0\int_{-q_\delta}^{q_\delta}{dq} \int_ {-\infty}^\infty 
 \frac{e^{i(q-q^\prime)k_{\rm F}x} dq^\prime} {(q^{2} - q^{\prime 2})+\delta 
\tilde a(q-\delta\nu q^\prime)},
\end{eqnarray}
in which the integral over $q^\prime $ extends over the entire real axis, and 
$I_{\delta\nu}^\epsilon$ corresponds to the undesirable singularities accounted 
within the extended limit of the integral over $q^\prime $. The integration of 
(\ref{Ir0})  over $q^\prime$ can be performed using Cauchy's integral theorem. 
For instance, after a cumbersome  integration over $q$~(see detail in  \ref{App:analytical}) we 
obtain  for $I_\RKKY$ (without the 
corrections),
\begin{eqnarray}\label{I_RKKY_NC}
 I_{\rm{RKKY}}^r&=&\pi I_{0}\Big\{\sin(\tilde{a}k_{\rm F}x)\Big[{{\rm Ci}} 
\left[(1-\tilde{a})2k_{\rm F} x\right]-
{{\rm Ci}}\left[(1+\tilde{a})2k_{\rm 
F}x\right] \Big]\nonumber \\
&&+\cos(\tilde{a}k_{\rm F}x)\Big[{\rm Si} 
\left[(1+\tilde{a})2k_{\rm F}x\right] 
-{{\rm Si}}\left[(\tilde{a} -1)2k_{\rm 
F}x\right]\nonumber \\
 && +{{\rm Si}}(2k_{\rm F}x)-{{\rm Si}}(-2k_{\rm F}x)\Big] -\ln\left|\frac{1-\tilde{a}}{1+\tilde{a}} 
\right|\sin(\tilde{a}k_{\rm F}x)  \Big\}.
\end{eqnarray}
In the equation above, ${\rm Si}(x)$ and ${\rm Ci}(x)$ are the known sine and 
cosine integral functions, respectively~\cite{footenote2}. To obtain the 
approximate expression we have to subtract the spurious contribution from 
the singularities. As an example, here we show in detail  the calculation of 
the correction for $I_{-+}^\epsilon$ to the integral 
$I_{-+}$~(see \ref{App:analytical}). Note that the unbalanced singularities occur when $q+q^\prime=0$ and $(q-q^\prime-\tilde{a})=0$ simultaneously, from which we find 
$q^\prime=-\tilde{a}/2=-q$. We can  evaluate the integral within an 
infinitesimal interval around this point as
\begin{eqnarray}\label{Imp_e}
I^{\epsilon}_{-+}=I_0e^{i\tilde{a}k_{\rm 
F}x}\int_{\frac{\tilde{a}}{2}-\epsilon}^{
\frac{\tilde{a}}{2}+\epsilon}
dq\int_{\frac{-\tilde{a}}{2}-\epsilon}^{-\frac{\tilde{a}}{2}+\epsilon}
\frac{dq^\prime}{(q+q^\prime)(q-q^\prime-\tilde a)},\nonumber \\
\end{eqnarray}
with $\epsilon \rightarrow 0$. On the rhs of the Eq.~(\ref{Imp_e}) we have 
already used that at the singular point under analysis,  $e^{i(q-q^\prime)k_{\rm 
F}x}=e^{i(\tilde{a} +\tilde{a})k_{\rm F}x/2}=e^{i\tilde{a}k_{\rm F}x}$. 
Performing the integral over $q^\prime $ we obtain
\begin{eqnarray}
I^{\epsilon}_{-+}=I_0e^{i\tilde{a}k_{\rm 
F}x} \!\int_{\frac{\tilde{a}}{2}-\epsilon}^{
\frac{\tilde{a}}{2}+\epsilon}\!\!
dq\frac{\ln|q-\frac{\tilde{a}}{2}+\epsilon|-\ln|q-\frac{\tilde{a}}{2}-\epsilon|}
{q-\frac{\tilde{a}}{2}}.\nonumber \\
\end{eqnarray}
After a simple change of variable $y=q-\tilde{a}/2$ we can write
$I_{-+}^{\epsilon}=I_0e^{i\tilde{a}k_{\rm 
F}x}\int_{-\epsilon}^{\epsilon}dy\left[\frac
{ \ln|y+\epsilon| }{y}-
 \frac{\ln|y-\epsilon|}{y}\right].
$
The integral here can be written in terms of the 
Dilogarithm function ${\rm Li}_{2}(x)=-\int_{0}^{x}dt\ln|1-t|/t$. With 
this  and  using $\rm{Li}_{2}(1)=\pi^{2}/6=-\rm{Li}_{2}(-1)$ 
\cite{Grobner} we can  write $I_{-+}^{\epsilon}\!\!=\!\!({\pi^2}/{2}) 
I_0e^{i\tilde{a}k_{\rm F}x}$. Proceeding likewise, we obtain the correction 
$I_{+-}^{\epsilon}= ({\pi^2}/{2})I_0e^{-i\tilde{a}k_{\rm F}x}$. Collecting 
all these terms, the correction for the RKKY coupling is given by 
$I^\epsilon_{\rm{RKKY}}=2\Re (I_{-+}^\epsilon+I_{+-}^{\epsilon *})= 4\pi 
I_{0}\left[\frac{\pi}{2} \cos(\tilde{a}k_{\rm F}x)\right].
$
% 
% \begin{eqnarray}
% I^\epsilon_{\rm{RKKY}}=2\Re (I_{-+}^\epsilon+I_{+-}^{\epsilon *})= 4\pi 
% I_{0}\left[\frac{\pi}{2} \cos(\tilde{a}k_{\rm F}x)\right].
% \end{eqnarray}
% 
This is the quantity we  must subtract from (\ref{I_RKKY_NC}) 
to obtain the approximated result. This result generalizes the correction found 
in Ref.~\cite{Zawadzki}. In the absence of SOC ($\tilde a =0$) 
 $I^\epsilon_{\rm{RKKY}}\!=\!2\pi^2 I_{0}$, which is exactly the 
correction discussed in Ref.~\cite{Zawadzki}.  The final expression for the 
RKKY coupling is then $I_\RKKY= I_{\RKKY}^r-I_\RKKY^\epsilon$. 

Carrying out similar calculations we obtain the analytical results for all 
inter-impurity couplings,
\begin{eqnarray}\label{Coupling_full_1_1}
 I_{\RKKY}&=&\pi 
I_{0}\Big\{\sin(\tilde{a}k_{\rm F}x)\Big[{\rm Ci}\left[\left(1-\tilde{a}
\right)2k_ { F } x\right]-  
{\rm Ci}\left[\left(1+\tilde{a}\right)2k_{\rm F}x\right]\Big] \nonumber \\
&&+\cos(\tilde{a}k_{\rm F}x)\Big[{\rm Si}\left[\left(1+\tilde{a}\right)2k_{\rm F}
x\right]-\rm{ {\rm Si} } \left[\left(\tilde { a }
-1\right)2k_{\rm F}x\right] \nonumber \\
&&+2{\rm Si}(2k_{\rm F}x)\Big]
 -{\ln}\left|\frac{1-\tilde{a}}{1+\tilde{a}}
\right|\sin(\tilde{a}k_{\rm F}x) 
 \Big\}-4\pi I_{0}\left[\frac{\pi}{2}\cos(\tilde{a}k_{\rm F}x)\right],\\
I_{\DM}&=&-\pi 
I_{0}\Big\{\cos(\tilde{a}k_{\rm F}x)\Big[{\rm Ci}\left[\left(1+\tilde{a}
\right)2k_ { F } x\right]-
 {\rm Ci}\left[\left(1-\tilde{a}\right)2k_{\rm F}x\right]\Big] \nonumber \\
&&+\sin(\tilde{a}k_{\rm F}x)\Big[{\rm Si}\left[\left(1+\tilde{a}\right)2k_{\rm F}
x\right]-\rm{ {\rm Si} } \left[\left(\tilde { a }
-1\right)2k_ {F}x\right]  \nonumber \\
&&+2{\rm Si}(2k_{\rm F}x)\Big]+{\ln}\left|\frac{1-\tilde{a}} {1+\tilde{a}} 
\right|\cos(\tilde{a}k_{\rm F}x)  \Big\}+4\pi I_{0}\left[\frac{\pi}{2} 
\sin\left(\tilde{a}k_{\rm F}x\right)\right],\\
 I_{\Is}&=&2\pi 
I_{0}\Big[{\rm Si}\left[\left(1+\tilde{a}\right)2k_{\rm F}x\right] 
-{\rm Si}\left[\left(\tilde{a}-1\right)2k_{\rm F}x\right]\Big] -4\pi 
I_{0}\left(\frac{\pi}{2}\right) 
-I_{\RKKY}.
\label{Coupling_full_1_3}
\end{eqnarray}

These rather complex expressions reduce to the known result $I_{\RKKY}=4\pi 
I_0[{{\rm Si}}(2k_{\rm F}x)-\pi/2]$  in the absence of SOC ($\tilde a=0$), 
that behaves as $\cos(2k_{\rm F}x)/x$ for large $x$ (with $I_\DM=I_\Is=0$). On 
the other hand, for $\tilde a \neq 0$ the leading terms for large  $x$ are
% 
%\begin{subequations}
 \begin{eqnarray}
I_{\RKKY}&=&-\pi 
I_{0}{\ln}\left|\frac{k_{\rm F}-2k_{\rm R}}{k_{\rm F}+ 2k_{\rm R}}
\right|\sin(2k_{\rm R}x),\label{asymptotic1}\label{asymptotic2}\\
I_{\DM}&=&-\pi 
I_{0}{\ln}\left|\frac{k_{\rm F}-2k_{\rm R}}{k_{\rm F}+ 2k_{\rm R}} 
\right|\cos(2k_{\rm R}x),\label{asymptotic2}\\
 I_{\Is}&=& \,\,\,\  \!\pi 
I_{0}{\ln}\left|\frac{k_{\rm F}-2k_{\rm R}}{k_{\rm F}+ 2k_{\rm R}}
\right|\sin(2k_{\rm R}x).\label{asymptotic3}
\end{eqnarray}
%\end{subequations}
% 
Here we have used $\tilde a\!=\!2k_{\rm R}/k_{\rm F}$. These remarkable
unsuppressed  oscillatory terms summarize the main result of our work. These 
terms contrast with the decaying behavior of the usual RKKY interaction in the 
absence of SOC. 

\begin{figure}[h]
\centering
\subfigure{\includegraphics[clip,width=4.5in]{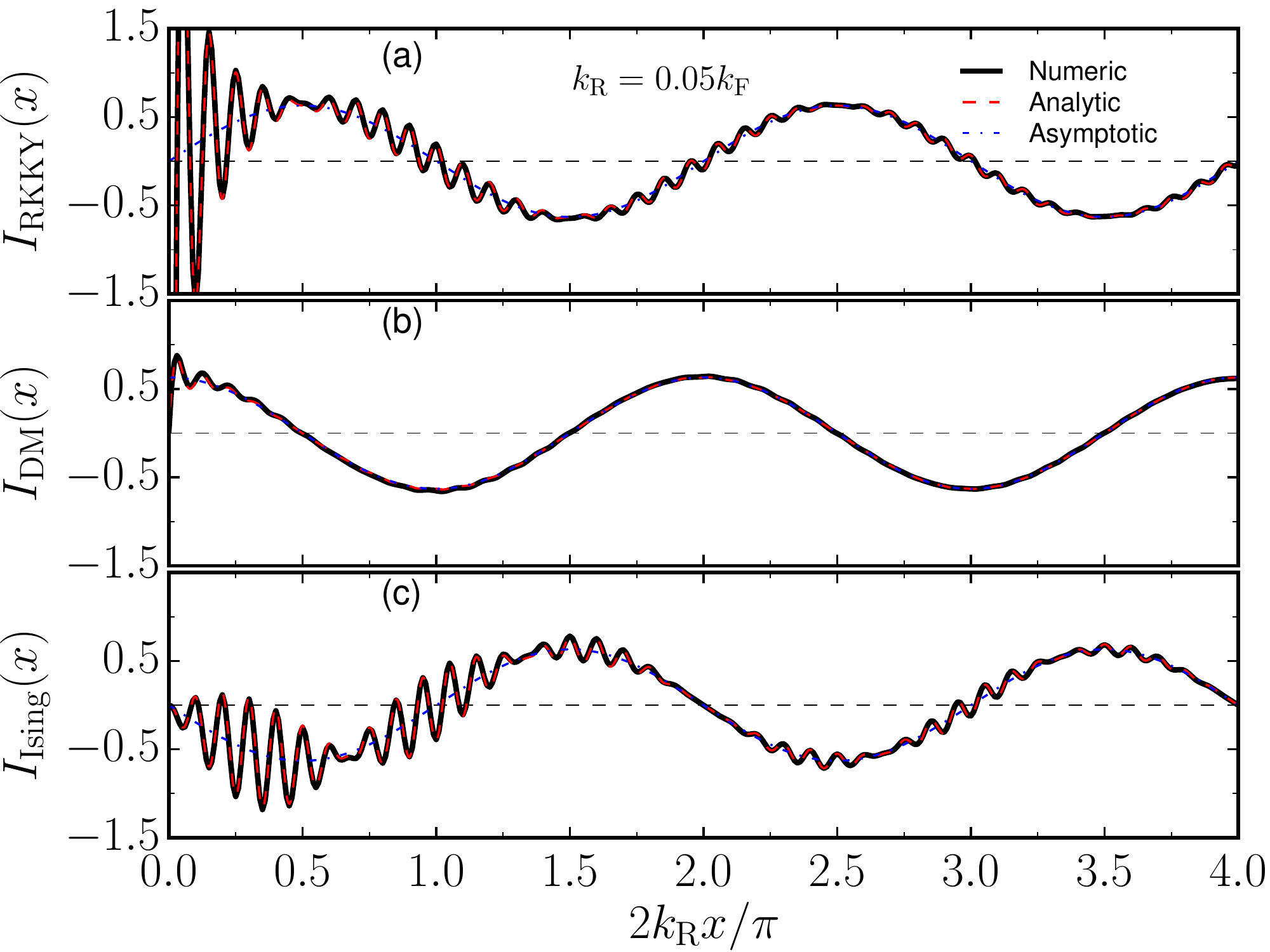}}
\caption{ RKKY (a)  Dzaloshinkii-Moryia (b) and Ising (c)
couplings as a function of the distance $x$ between the impurities. 
Solid black lines correspond to the results obtained by direct numerical integration 
of Eq.~(\ref{coupling_integral0}), dashed red lines correspond to the analytical results of
Eqs.~(\ref{Coupling_full_1_1})-(\ref{Coupling_full_1_3}) and dash-dot blue lines show the asymptotic behavior of the coupling
given by the Eqs.~(\ref{asymptotic1})-(\ref{asymptotic3}).} 
\label{fig2}
\end{figure}

Before discussing these results, we compare the analytical results of 
Eqs.~(\ref{Coupling_full_1_1})-(\ref{Coupling_full_1_3}) with the ones obtained by direct numerical integration 
of (\ref{coupling_integral_1}) for $\tilde a=0.1$ ($k_{\rm R}=0.05 k_{\rm 
F}$). The results are shown in Fig.~\ref{fig2}. Panels \ref{fig2}(a), 
\ref{fig2}(b) and \ref{fig2}(c) correspond to the $I_\RKKY$, $I_\DM$ and 
$I_\Is$, respectively. Dashed red lines correspond to the  analytical results of Eqs.~(\ref{Coupling_full_1_1})-(\ref{Coupling_full_1_3}) while solid black lines  refer to the numerical results obtained by direct integration of  Eq.~(\ref{coupling_integral_1}). The dash-dot  blue lines show the asymptotic behavior of the coupling given by the  Eqs.~(\ref{asymptotic1})-(\ref{asymptotic3}). The extraordinary agreement between our numerical and analytical results shown in Fig.~\ref{fig2} confirms that we have indeed obtained very good approximate expressions for all couplings.  

The striking features are the undamped slow oscillations in the coupling due to the SOC mentioned earlier. The fast oscillations along the slow oscillating line is the traditional behavior of the RKKY interaction and result from the polarization of the Fermi sea by one impurity and ``felt'' by the other one. They are described by the ${\rm Ci}$ and ${{\rm Si}}$ functions of Eqs.~(\ref{Coupling_full_1_1})-(\ref{Coupling_full_1_3}) and 
have the traditional period $2\pi /2k_{\rm F}$. Note also in 
Eqs.~(\ref{Coupling_full_1_1})-(\ref{Coupling_full_1_3}) the extra terms $\pm 2\tilde a k_{\rm F}x=\pm 2 k_{\rm R}x$ in the arguments of the functions ${{\rm Ci}}$ and ${\rm {\rm Si}}$. They are responsible for the curious  beating patterns observed in the fast oscillations.     
Physically, the beating patterns can be understood in the following way: the original RKKY interaction exhibits an oscillation with frequency of $2k_F$. In the absence of SOC, the spin up and down bands are degenerated leading to a single Fermi momenta $k_F$. Here, on the other hand, electrons move freely in different helical bands possessing Fermi momenta are slightly shifted as compared to each other. Since the real spin basis representation is a linear combination of each Rashba bands, the resulting spin polarization is a combination of two oscillating terms whose phase are slightly shifted. This renders the beating pattern observed along the distance between the impurities. These beating patterns are akin to what was found in~\cite{Valizadeh} for the RKKY interaction in spin-polarized bands.

\section{Discussions}
The unsuppressed oscillations obtained here can be understood as follows: 
after a given  electron is scattered by the first impurity it travels  
throughout the quantum wire while its spin precesses due to the SOC. Since 
the momentum and spin are coupled together, this precession continues 
coherent until it collides with the second impurity. Somehow, the  
momentum-spin lock produced by the SOC in this 1D system  provides a  natural 
protection (not topological) that prevents the  suppression of the couplings. 
To provide a better physical intuition, let us analyze the scattering processes 
involved in the second order perturbation theory. If we write the Hamiltonian 
(\ref{H_Kondo}) onto the Rashba basis we obtain 
\begin{eqnarray}\label{H_Kondo}
H_i&=&\frac{J}{N}\sum_{kk^\prime}e^{-i(k-k^\prime)x_i}\left[
\tilde S^z_i\left(c^\+_{ k^\prime+ } c_ { k -} -c^\+_ { 
k^\prime -}c_{k\+}\right)+ \tilde S^+_ic^\+_{k^\prime\-}c_{k+}  + \tilde S^-_ic^+_{k^\prime +}c_{k-}\right].
\end{eqnarray}
Here, again, the tilde on the spin operators emphasizes that they  are also 
written on the Rashba basis, meaning that the ``spin'' scattering processes 
correspond to removing electrons from one band to another. 
Although the processes formally very much similar to those ones that occur in 
the absence of the SOC. Here,  by virtue of the shift property 
$\varepsilon_+(k)=\varepsilon_-(k+Q)$ (where $Q=2k_{\rm R}$), the  
spin-flip processes  in the second order perturbation theory involve 
intermediate states  whose momenta is separated from the initial states by 
$2k_{\rm R}$ (for forward scattering process) or $2k_{\rm F}$ 
(for backscattering processes). In this sense, at zero temperature, conserving 
momenta scatterings are prevented by the SOC, which inhibits the  decay of the 
couplings in the system. 
This analysis also allows us to understand  enhancement of the oscillation 
amplitudes of Eqs.~(\ref{asymptotic1})-(\ref{asymptotic3}). When $k_{\rm F}\longrightarrow 2k_{\rm 
R}$  the Fermi momenta matches precisely the distance (in the momentum space) 
between the two bands, providing a resonant forward scattering. In 
reality, similarly to all the traditional approach to RKKY interaction, our 
results are limited distances smaller than the coherent length of the 
material. For distances larger than this characteristic length, other 
scattering processes have to be taken into account in the conduction electron 
propagation.

Somewhat similar to our results was found by J. Simonin~\cite{Simonin}. He has 
found a spin-spin correlation between two magnetic moments induced by 
spin that extends also to distances longer than those of the traditional RKKY 
interaction.  Our results also resemble the persistent spin 
helix~\cite{Bernevig,Awschalom} in which a ``right'' combination 
of Rahsba and Dresselhaus SOCs produces a long lived spin excitation in the 
system. 
This contrasts with the traditional scattering in the absence of the SOC, in 
which there is a $Q=0$ scattering processes is allowed since the 
$\varepsilon_\up (k)=\varepsilon_\down(k)$.
Previous studies usually employ a very attractive expression based on 
real space Green's functions~\cite{Imamura}. However, as thoroughly discussed  
by Valizadeh~\cite{Valizadeh} the expression should be avoided in 1D systems. 
Essentially, the reason is because in the  derivation of the equation (5) of 
Ref.~\cite{Imamura} there is a change in the order of integration in double 
integral that should not be made in one-dimension. Here we circumvent this problem  by directly performing the integrals (\ref{coupling_integral_1}) both 
analytically and numerically. See detailed discussion in \ref{App:Greens_function}.

To interpret our results, let us recall that the mechanism  responsible for the decaying oscillations in the RKKY interaction results from the existence of a Fermi sea.  Under the second order perturbation theory perspective, the propagating electrons with momentum $k_F$ suffer scatterings with the Fermi sea. In the absence of SOC, these scatterings occur independently of the spin orientation of the  propagating electrons. In contrast, in the presence of SOC, spin and momentum are locked together. As a result, an scattering  can only occur if the spin orientation is modified accordingly. In 1D, backward scattering, for instance, has to be accompanied by a spin flip. Therefore, some of scatterings allowed in the absence of SOC are prevented when spin and momentum are coupled together.

\section{Conclusions}
We have investigated the exchange interaction 
between two magnetic impurities  mediated by conduction electrons in a 
one-dimensional system with SOC. We revisited the calculation 
for the RKKY interaction in one-dimensional system by employing a 
straightforward second order perturbation theory of a two-impurities Kondo 
model. We find that in the presence of the SOC, the known  RKKY, 
Dzaloshinkii-Moryia and Ising exchange interactions exhibit an additional 
oscillation resulting from the spin precession of the conduction electrons that 
mediate the exchange interactions. More interestingly, these additional 
oscillations  do not decay with distance between the impurities. This is in 
sharp contrast to the results in the absence of the SOC that shows an RKKY 
coupling the behaves as $x^{-1}$. Moreover, our 
results also contrast with the recent calculations of RKKY interaction in
1D system with spin orbit~\cite{Imamura}. The apparent difference between our 
results and the those from~\cite{Imamura} arises from the fact that the expression used 
in the later cannot be straightforwardly applied in the 1D systems~\cite{Valizadeh}, specially
in the presence of SOC Here, we avoid the problem by  performing explicitly the integral 
resulting from the second order perturbations theory. Our work extends the expression for the 
1D indirect exchange interactions to the case in which SOC is present. This is not only important
because it is fundamentally distinct from the usual case in the absence of the SOC but also may
be useful for practical application where long-distance couplings are relevant. Magnetic impurities
in materials such as GaAs/AlGaAs~\cite{Picciott} or InAs~\cite{Gao} spin-orbit coupled quantum
wires are examples of potential candidates for experimental verification of our predictions.
\ack

We thank  Professors G. Ferreira, M. A. Boselli, G. B. Martins and 
E. V. Anda for great discussions. We also acknowledge 
financial support from CNPq, CAPES and FAPEMIG.  
\appendix
\section*{Appendix}
\section{Effective inter-impurities Hamiltonian}
\label{App:Perturbation}
%  \noindent
To derive the effective inter-impurities Hamiltonian we follow the traditional 
approach use to obtain the usual RKKY.  We assume 
\begin{equation}\label{H_RD}
H_{\wire}= \sum_k \left(\varepsilon_k\delta_{ss^\prime}
-\alpha_R k\sigma^{y}_{s s^\prime}\right) c^\dagger_ { k s }c_{k s^\prime},
\end{equation}
as the unperturbed Hamiltonian that includes the spin-orbit interaction. The  
perturbation 
 \begin{eqnarray}\label{Hi}
H_i&=&\frac{J}{N}\sum_{kk^\prime}e^{-i(k-k^\prime)
x_i}\left[S^z_i\left(c^\+_ { k^\prime\up } c_ { k\up } -c^\+_ { 
k^\prime\dn}c_{k\dn}\right) + S^+_ic^\+_{k^\prime\dn}c_{k\up} 
+ S^-_ic^+_{k^\prime\up}c_{k\dn}\right]
\end{eqnarray}
accounts for the impurities.  To apply the second order perturbation theory we 
diagonalize the Hamiltonian (\ref{H_RD}). This is achieved by defining the 
new operators by the transformation 
 \begin{equation}\label{transformation}
  \left(\begin{array}{c}
         c_{k+} \\
         c_{k-}
        \end{array}\right)=\mathcal{U}_{k} \left(\begin{array}{c}
         c_{k\up} \\
         c_{k\down}
        \end{array}\right),
 \end{equation}
where
\begin{equation}\label{matrix}
 \mathcal{U}_{k}=\frac{1}{\sqrt{2}}\left(\begin{array}{cc}
                                      1 & 1 \\
                                      -i & i
                                     \end{array}
\right),
\end{equation}
is a unitary matrix. The transformation above 
corresponds to a momentum-dependent rotation in  the spin space. 
In the new base $H_{\wire}$ acquires the diagonal form
\begin{eqnarray}
\tilde H_{\wire}=\sum_{k h} \varepsilon_{kh}c^ 
\dag_{kh}c_{kh},
\end{eqnarray}
in which $h=+,-$ is the helical quantum number and $\varepsilon_{k h}=\hbar 
k^2/2m^*+h\alpha_R k$ are the eigenvalues of $H_{\wire}$. The eigenstates are 
then defined as $|k,h \rangle$  such that $H_\wire|k,h \rangle=\varepsilon_{k 
h}|k,h \rangle$. 

For simplicity, here we assume  impurities have spin $1/2$ so that the spin 
operators can be easily written in terms of fermion operators as 
$S^{z}=(d^\+_{\up}d_{\up}-d^\+_{\down}d_{\down})/2$, 
$S^{+}=d_{\up}^{\dag}d_{\down}$, and $S^{-}=d_{\down}^{\dag}d_{\up}$, where 
$d^\+_s$ ($d_s$) corresponds to the creation (annihilation) spin-1/2 fermion 
operator.  This is very useful because we can now perform the same rotation 
(\ref{transformation}) for these fermion operators, after which we can rewrite 
(\ref{Hi}) as
\begin{eqnarray}\label{impurit}
H_i&=&\frac{J}{N}\sum_{kk^\prime}e^{-i(k-k^\prime)
x_i}\left[\tilde{S}_{i}^z\left(c^\+_ { k^\prime + } c_ { k + } -c^\+_ { 
k^\prime -}c_{k -}\right)+ \tilde{S}_{i}^+c^\+_{k^\prime -}c_{k+} 
+ \tilde{S}_{i}^-c^+_{k^\prime +}c_{k-}\right].
\end{eqnarray}
Here, the $\tilde{S}_{i}^{z}$ emphasizes that the impurity spin 
operators are written on the rotated spin basis.
Having the eigenstates and eigenergies of the unperturbed Hamiltonian, the 
prescription to obtain the RKKY coupling is to compute  the correction to the 
total energies up to the second order perturbation theory. To account for the 
degrees of freedom of the impurities, an eigenstate of $H_\wire$ can be written 
as $|k,h\rangle$, where $h$ is the helical quantum number.

The textbook 
expression for the second order energy correction  can be written as
% 
%\begin{strip}
\begin{equation}\label{pertu}
H_{\rm eff}=\sum_{k,h}^{\rm occ.}\, \sum_{{k^\prime, h^\prime}}^{\rm 
empty}\frac{ \langle k,h|(H_1+H_2)|k^\prime, h^\prime\rangle\langle 
k',h'|(H_1+H_2)|k,h\rangle}{\varepsilon_{kh}-
 \varepsilon_{k'h'}},
\end{equation}
%\end{strip}
% 
where $H_{i}$ is given by (\ref{impurit}). In the Eq.~(\ref{pertu})  we 
assume that we are at temperature $T=0$, in  which case, the bands are fully 
occupied up to the Fermi level while fully empty above it. The exchange energy 
is only due the mixed terms of (\ref{pertu}), we thus drop the 
self-interaction terms and write
\begin{eqnarray}\label{mixed}
\!\!\!\!\! \!\!\!\!\!\!\!\!\! H_{\rm eff}&=&\sum_{k,h}^{\rm occ.}\, \sum_{{k^\prime, 
h^\prime}}^{\rm 
empty}\left[\frac{ \langle k,h|H_1|k^\prime, h^\prime\rangle\langle 
k',h'|H_2|k,h\rangle}{\varepsilon_{kh}-
 \varepsilon_{k'h'}}+\frac{ \langle k,h|H_2|k^\prime, h^\prime\rangle\langle 
k',h'|H_1|k,h\rangle}{\varepsilon_{kh}-
 \varepsilon_{k'h'}}\right].
 \end{eqnarray} 
The non-vanish contributions of (\ref{mixed}) can be calculated applying the creator and annihilator operators on the state $|kh\rangle$. For example 
$\tilde{S}^{z}c_{k'+}^{\dag}c_{k+}|k,+\rangle=\tilde{S}^{z}|k',+\rangle$, 
$\tilde{S}^{+}c_{k'-}^{\dag}c_{k+}|k,+\rangle=\tilde{S}^{+}|k',-\rangle$,
$\tilde{S}^{-}c_{k'+}^{\dag}c_{k-}|k,-\rangle=\tilde{S}^{-}|k',+\rangle$. 
Using these relations we obtain
\begin{eqnarray}\label{NVT}
 \langle k,+|H_{1}|k',+\rangle&=&\frac{J}{N}\sum_{kk'}e^{i(k-k')x_{1}}
\tilde{S}_{1}^{z}; \\
 \langle k,-|H_{1}|k',-\rangle&=&-\frac{J}{N}\sum_{kk'}e^{i(k-k')x_{1}}
 \tilde{S}_{1}^{z}, \\
 \langle k,-|H_{1}|k',+\rangle &=&\frac{J}{N}
 \sum_{kk'}e^{i(k-k')x_{1}}\tilde{S}_{1}^{+};\\ 
 \langle k+|H_{1}|k',-\rangle &=&\frac{J}{N}
 \sum_{kk'}e^{i(k-k')x_{1}}\tilde{S}_{1}^{-}.
 \end{eqnarray}
Here we have used the orthogonality relation $\langle 
k,h|k',h'\rangle=\delta_{kk'}\delta_{hh'}$. Carrying out the calculation for
$\langle k,h|H_{2}|k',h'\rangle$ we obtain  similar results. Unlike the usual case of absence of SOC, in which the energy is equal for 
both spin components, here the energies  $\varepsilon_{kh}$ 
depend of the helical number.  Using $\varepsilon_{k h}=\hbar 
k^2/2m^*+h\alpha_R k$, the energy differences that appears in the denominator of 
the four non-vanishing terms of (\ref{mixed}) 
are
\begin{eqnarray}\label{De}
%\begin{array}{r@{}l}
\varepsilon_{k+}-\varepsilon_{k'+}&=&\frac{\hbar^{2}}{2m^*}(k^{2}-k'^{2}
)+\alpha_R(k-k');\\
\varepsilon_{k-}-\varepsilon_{k'-}&=&\frac{\hbar^{2}}{2m^*}(k^{2}-k'^{2}
)-\alpha_R(k-k'), \\
 \varepsilon_{k-}-\varepsilon_{k'+}&=&\frac{\hbar^{2}}{2m^*}(k^{2}-k'^{2}
)-\alpha_R(k+k'); \\
\varepsilon_{k+}-\varepsilon_{k'-}&=&\frac{\hbar^{2}}{2m^*}(k^{2}-k'^{2} 
)+\alpha_R(k+k').
%\end{array}
 \end{eqnarray}
Replacing the results of the Eqs.~(\ref{NVT}-A.12) and Eqs.~(\ref{De}-A.16) into Eq.~(\ref{mixed}) we obtain

\begin{eqnarray}\label{H_eff}
H_{\rm 
imp}&=&(I_{++}+I_{++}^*+I_{--}+I_{--}^*)\tilde{S}^{z}_{1}\tilde{S}^{z}_{2}+(I_{
-+
}+I^*_{+-})\tilde{ S}^{+}_{1}\tilde{S}^{-}_{2} 
  +(I_{+-}+I^*_{-+})\tilde{S}^{-}_{1}\tilde{S}^{+}_{2},\nonumber\\
\end{eqnarray}
with
\begin{eqnarray}\label{coupling_integral_sum}
I_{\delta\nu}\!=\!\frac{J^2}{N^2}\sum_k^{\rm occupied}\, \sum_{k^\prime}^{\rm 
empty}\frac{e^{i(k-k^\prime)x}} {\frac{\hbar^{2}}{2m^*}(k^{2} - 
k^{^\prime 2})+\delta\alpha_{\rm R}(k-\delta\nu k^\prime)},
\end{eqnarray}
in which  $x=x_2-x_1$ is the distance between the impurities. The effective 
Hamiltonian (\ref{H_eff}) can be written in a more compact form
\begin{eqnarray}\label{eq1}
\tilde H_{\rm{imp}}& =&I_\parallel \tilde{S}^{z}_{1}\tilde{S}^{z}_{2} + 
I_{\perp}\tilde{S}^{+}_{1}\tilde{S}^{-}_{2} + 
I_{\perp}^*\tilde{S}^{-}_{2} \tilde{S}^{+}_{1},
\end{eqnarray}
where we have defined $I_{\parallel}=2\Re (I_{++}+I_{--})$ e 
$I_{\perp}=(I_{-+}+I_{+-}^*)$. We now transform the summations into integrals 
using the usual prescription $(1/N)\sum_k \rightarrow (1/2\pi\int dk)$ in the 
limit $N\rightarrow \infty$, so that the Eq.~(\ref{coupling_integral_sum}) can 
now be written as
\begin{eqnarray}\label{coupling_integral}
I_{\delta\nu}={J^2}\int_{-k_\delta}^{k_\delta}\frac{dk}{2\pi}\int_{
|k^\prime|>k_\delta } \frac{dk^\prime}{2\pi} \frac{e^{i(k-k^\prime)x} {
} } {\frac{\hbar^{2}}{2m^*}(k^{2} - 
k^{^\prime 2})+\delta\alpha_{\rm R}(k-\delta\nu k^\prime)},
\end{eqnarray}
Here we also used the fact that, because of the SOC, the bands $+$ and $-$ have 
different Fermi momenta, namely $k_\delta=k_F+\delta k_{\rm R}$ 
(for $\delta=+,-$). In the helical basis, the Hamiltonian (\ref{eq1}) has the 
form of a anisotropic Heisenberg Hamiltonian. Although simple, it hides the 
physics we want to study here. We can rewrite the impurity  operators  on the 
reals  spin basis, on which we have 
\begin{eqnarray}
\tilde{S}^{+}_{1}\tilde{S}^{-}_{2} &=& \mathbf{S}_{1}\cdot 
\mathbf{S}_{2}+i(\mathbf{S} _ { 1 }\times 
\mathbf{S}_{2})\cdot \hat{\bf y}-S^{y}_{1}S^{y}_{2},\\
\tilde{S}^{-}_{1}\tilde{S}^{+}_{2}&=&\mathbf{S}_{1} 
\cdot \mathbf{S}_{2}-i(\mathbf{S}_{1} \times 
\mathbf{S}_{2})\cdot \hat{\bf y}-S^{y}_{1}S^{y}_{2}, \\
\tilde{S}^{z}_{1}\tilde{S}^{z}_{2}&=& S^{y}_{1}S^{y}_{2}.
\end{eqnarray}
Thus, in the real spin space,  the exchange Hamiltonian is given by  
\begin{eqnarray}\label{Exchange}
\tilde H_{\rm{imp}}=I_{\rm{RKKY}}\mathbf{S}_{1}\cdot
\mathbf{S}_{2}+I_{\rm{DM}}[(\mathbf{S}_{1}\times \mathbf{S}_{2})
\cdot\hat{\bf y}] +I_{\rm{Ising}}S^{y}_{1}S^{y}_{2},
\end{eqnarray}
where, $I_{\rm{RKKY}}=2\Re [I_{\perp}]$,  $I_{\rm{DM}}=-2\Im 
[I_{\perp}]$ and $I_{\rm{Ising}}= I_{\parallel}-2\Re [I_{\perp}]$  are the 
known RKKY, Dzaloshinkii-Moryia, and the Ising couplings. 
 
\section{Analytical calculation of the couplings} \label{App:analytical}
 \noindent 
We now focus on the calculation of the couplings $I_\RKKY$, $I_\DM$  and 
$I_\Is$. This requires performing the integrals (\ref{coupling_integral}). To 
simplify the notation we define the dimensionless variables $q=k/k_{\rm F}$, 
$q^\prime=k^\prime/k_{\rm F}$, together with $a=2k_{\rm R}$, with  $k_{\rm 
R}=m\alpha_{\rm R}/\hbar^{2}$, and $\tilde{a}=a/k_{\rm F}$. With these 
definitions the 
Eq.~(\ref{coupling_integral}) acquires the form
\begin{eqnarray}\label{adcoupling}
I_{\delta\nu}=I_{0}\int_{-q_{\delta}}^{q_{\delta}}dq\int_{
|q^\prime|>q_{\delta}}
dq^\prime\frac{e^{i(q-q^\prime)k_{F}x}}{(q^{2} - 
q^{^\prime 2})+\delta\tilde{a}(q-\delta\nu q^\prime)},
\end{eqnarray}  
where $I_{0}=J^{2}m/2\pi^{2}\hbar^{2}$, and $q_{\delta}=1+\delta \tilde{a}/2$. 
An important point here that should be highlighted is that the order of the 
integrations above should not be changed as discussed by Yafet\cite{Yafet}. 
Later  Valizedeh~\cite{Valizadeh} revisited the problem and noted that the 
problem is that the integrals \ref{coupling_integral} do not obey the 
Fubini's condition~\cite{Fubini,Friedman}, leading to different results 
depending on the order in which the integrations are performed.  
Here we keep the order of integrations as it is in 
Eq.~(\ref{coupling_integral}), avoiding the aforementioned problem. To perform 
the integral over $q$ using the residues theorem we need to extend it to the 
entire real axis. With this we can write
\begin{eqnarray}\label{Ir}
I_{\delta\nu}^r=I_0\int_{-q_\delta}^{q_\delta}{dq} \int_ {-\infty}^\infty 
dq^\prime \frac{e^{i(q-q^\prime)k_Fx}} {(q^{2} - q^{\prime 2})+\delta 
\tilde a(q-\delta\nu q^\prime)}.
\end{eqnarray}
This deformation of the integral limits introduce undesirable 
contributions. If we are able to account for these extra contributions 
separately, we can subtract them from the final results to obtain the correct 
expression. In the absence of SOC, the integrand of (\ref{Ir}) is antisymmetric 
under the exchange $q \longleftrightarrow q^\prime$, thus the extra 
contributions added to the results are solely those 
coming from corresponds to the singularities  occurring at $q=q^\prime$. 
However,  in the presence of the SOC ($\tilde a \neq 0$) the integrand is no 
longer antisymmetric. Therefore, there are  contributions other than those 
arising from the singularities. Here we assume that the only relevant additional contributions are those  arising from the singularities of 
\ref{Ir}. Thus, within this approximation,  we can write  
$I_{\delta\nu}=I^{r}_{\delta\nu} -I^{\epsilon}_{\delta\nu}$, where and 
$I^{\epsilon}_{\delta\nu}$ corresponds to 
the undesirable singularities. We first integrate over $q^\prime$ and then over 
$q$. The Eq.~(\ref{Ir}) can be written as 
\begin{equation}
 I_{\delta\nu}^r=I_0\int_{-q_\delta}^{q_\delta}{dq}I_{\delta\nu}^{rq},
\end{equation}
where we have defined  
\begin{equation}
 I_{\delta\nu}^{rq}=\mathcal{P}\oint 
dz \frac{e^{i(q-z)k_Fx}} {(q^{2} - z^{2})+\delta \tilde a(q-\delta\nu z)}.
\end{equation}
in the above $\mathcal{P}[\cdots]$ denote the  Cauchy principal value. 

Let us start with by calculating $I_{++}^{rq}$ 
that has the form 
\begin{eqnarray}
I^{rq}_{++}=\mathcal{P}\oint 
dz\frac{e^{i(q-z)k_{F}x}}{q^{2}-z^{2}+\tilde a(q-z)}.
\end{eqnarray}
Closing the contour on the lower half-plane and using the residues theorem we 
obtain
\begin{equation}\label{++}
 \mathcal{P}[I^{rq}_{++}]=\frac{2\pi}{2q+\tilde{a}}\sin[(q+\tilde{a}/2)k_{F}x] 
e^{i(q+\tilde{a}/2)k_{F}x}.
\end{equation}
Noticing from (\ref{Ir}) that we can obtain $I_{--}^{rq}$ by doing 
$\tilde{a}\rightarrow -\tilde{a}$ in the Eq.~(\ref{++}). Therefore we 
immediately  obtain
\begin{equation}\label{--}
 \mathcal{P}[I^{rq}_{--}]=\frac{2\pi}{2q-\tilde{a}}\sin[(q-\tilde{a}/2)k_{F}x] 
e^{i(q-\tilde{a}/2)k_{F}x}.
\end{equation}
Proceeding in a similar way for the other two integrals we obtain
\begin{equation}\label{+-}
 \mathcal{P}[I^{rq}_{+-}]=\frac{2\pi}{2q+\tilde{a}}
 \sin[(q+\tilde{a}/2)k_{F}x]e^{i(q-\tilde{a}/2)k_{F}x}.
\end{equation}
and
\begin{equation}\label{-+}
 \mathcal{P}[I^{rq}_{-+}]=\frac{2\pi}{2q-\tilde{a}}
 \sin[(q-\tilde{a}/2)k_{F}x]e^{i(q+\tilde{a}/2)k_{F}x}.
\end{equation}
Collecting the results (\ref{++})-(\ref{-+}) and grouping them properly, we 
obtain 
\begin{eqnarray}\label{IRKKY_r}
 I^{r}_{\rm{RKKY}}&=&2\Re(I_{-+}^{r}+I_{+-*}^{r})\nonumber \\
 &=&4\pi I_{0} 
\left[\int_{-q_{+}}^{q_{+}} dq\frac{\cos[(q-\tilde{a}/2) 
k_{F}x]\sin[(q+\tilde{a}/2) k_{F}x]}{2q+\tilde{a}} 
\right. \nonumber \\
 &&\left.+
 \int_{-q_{-}}^{q_{-}} dq\frac{\cos[(q+\tilde{a}/2)k_{F}x] 
\sin[(q-\tilde{a}/2)k_{F}x]}{2q-\tilde{a}} \right],\nonumber \\
\end{eqnarray}
\begin{eqnarray}
\label{IDM_r}
 I^{r}_{\rm{DM}}&=&-2\Im(I_{-+}^{r}+I_{+-*}^{r})\nonumber \\
 &=&-4\pi I_{0} 
\left[\int_{-q_{-}}^{q_{-}} dq\frac{\sin[(q+\tilde{a}/2)k_{F}x] 
\sin[(q-\tilde{a}/2)k_{F}x]}{2q-\tilde{a}}\right. \nonumber \\
 &&\left.-  \int_{-q_{+}}^{q_{+}} dq\frac{\sin[(q+\tilde{a}/2)k_{F}x] 
\sin[(q-\tilde{a}/2)k_{F}x]}{2q+\tilde{a}}\right],\nonumber \\
\end{eqnarray}
and 
\begin{eqnarray}\label{IIsing_r}
  I^{r}_{\rm{Ising}}&=&2\Re(I_{++}^{r}+I_{--}^{r})-I^{r}_{\rm{RKKY}}\nonumber \\
  &=&2\pi I_{0}\left[\int_{-q_{+}}^{q_{+}} dq\frac{\sin[(2q+\tilde{a}) 
k_{F}x]}{2q+\tilde{a}} +\int_{-q_{-}}^{q_{-}} dq\frac{\sin[(2q-\tilde{a}) 
k_{F}x]}{2q -\tilde{a}}\right]-I^{r}_{\rm{RKKY}}.\nonumber \\
 \end{eqnarray}
The superindices ``$r$'' denote the uncorrected results, i.e., before 
subtracting the extra contribution.  
The six  integrals appearing in the expressions 
(\ref{IRKKY_r})-(\ref{IIsing_r}) above are rather complicated but can still 
be computed  analytically. After a tiresome work, apart from additive 
constants,  we obtain the expressions for the undefined  integrals 
\begin{eqnarray}\label{Int1}
 \int   
dq\frac{\sin[(2q+\tilde{a})k_{F}x]}{2q+\tilde{a}}&=&\frac{1}{2}{\rm Si}[
(2q+\tilde{a})k_{F}x],
\end{eqnarray}
\begin{eqnarray}
 \int dq\frac{\sin[(2q-\tilde{a})k_{F}x]}{2q-\tilde{a}}&=&\frac{1}{2}{\rm Si}[
(2q-\tilde{a})k_{F}x], 
\end{eqnarray}
% 
%\begin{strip}
\begin{eqnarray}
 \int  
 dq\frac{\cos[(q-\tilde{a}/2)k_{F}x]\sin[(q+\tilde{a}/2)k_{F}x]}{2q+\tilde{a}}&=&
 \frac{1}{4}\Big\{-\sin(\tilde{a}k_{F}x){\rm Ci}[(2q+\tilde{a})k_{F}x]. \nonumber \\ 
&&+\cos(\tilde{a}k_{F}x)
 {\rm Si}[(2q+\tilde{a})k_{F}x] \nonumber \\
 &&+\sin(\tilde{a}k_{F}x)\ln[2(2q+\tilde{a})]
\Big\},\nonumber \\
\end{eqnarray}
\begin{eqnarray}
 \int  
 dq\frac{\cos[(q+\tilde{a}/2)k_{F}x]\sin[(q-\tilde{a}/2)k_{F}x]}{2q- 
\tilde{a}}&=& \frac{1}{4}\Big\{\sin(\tilde{a}k_{F}x){\rm Ci}[(2q- 
\tilde{a})k_{F}x]\nonumber \\
&&-\cos(\tilde{a}k_{F}x) {\rm Si}[(\tilde{a}-2q) 
k_{F}x]\nonumber \\ 
&&-\sin(\tilde{a}k_{F}x)\ln[2(\tilde{a}-2q)]\Big\},\nonumber \\
\end{eqnarray}
\begin{eqnarray}
 \int  
 dq\frac{\sin[(q-\tilde{a}/2)k_{F}x]\sin[(q+\tilde{a}/2)k_{F}x]}{2q 
+\tilde{a}}&=& \frac{1}{4}\Big\{-\cos(\tilde{a}k_{F}x) 
{\rm Ci}[(2q+\tilde{a})k_{F}x] \nonumber \\
&&-\sin(\tilde{a}k_{F}x)
 {\rm Si}[(2q+\tilde{a})k_{F}x] \nonumber \\
 &&+\cos(\tilde{a}k_{F}x){\ln}[2(2q+\tilde{a})]\Big\},\nonumber \\
\end{eqnarray}
and 
\begin{eqnarray}\label{Int6}
 \int  
 dq\frac{\sin[(q+\tilde{a}/2)k_{F}x]\sin[(q-\tilde{a}/2)k_{F}x]}{2q- 
\tilde{a}}&=& \frac{1}{4}\Big\{-\cos(\tilde{a}k_{F}x){\rm Ci}[(2q-\tilde{a}) 
k_{F}x] \nonumber \\
&&-\sin(\tilde{a}k_{F}x) {\rm Si}[(\tilde{a}-2q)k_{F}x] \nonumber \\
 &&\quad+\cos(\tilde{a}k_{F}x)\ln[2(\tilde{a}-2q)]\Big\}.\nonumber \\
\end{eqnarray}
In the above we use the usual definitions 
\begin{eqnarray}
{\rm Ci}(x)&=&\int_0^x\frac{\cos( t)}{t} dt \\
{\rm {\rm Si}}(x)&=&\int_0^x\frac{\sin( t)}{t} dt.
\end{eqnarray}
After imposing the proper limits to the results (\ref{Int1})-(\ref{Int6})  and 
some algebraic manipulations we can write 
\begin{eqnarray}\label{Coupling_full}
I^{r}_{\RKKY}&=&\pi 
I_{0}\Big\{\sin(\tilde{a}k_{F}x)\Big[{\rm Ci}\left[(1-\tilde{a})2k_{F} 
x\right]-  {\rm Ci}\left[(1+\tilde{a})2k_{F}x\right]\Big] \nonumber \\
&&+\cos(\tilde{a}k_{F}x) 
\Big[{\rm Si}\left[(1+\tilde{a})2k_{F}x\right]-{\rm Si}\left[(\tilde{a} 
-1)2k_{F}x\right] \nonumber \\ 
&&+2{\rm Si}(2k_{F}x)\Big]-{\ln}\left|\frac{1-\tilde{a}}{1+\tilde{a}} \right|\sin(\tilde{a}k_{F}x) 
 \Big\},\\
I^{r}_{\DM}&=&-\pi  I_{0}\Big\{\cos(\tilde{a}k_{F}x)\Big[{\rm Ci} 
\left[(1+\tilde{a})2k_{F}  x\right]-  
{\rm Ci}\left[(1-\tilde{a})2k_{F}x\right]\Big] \nonumber \\
&&+\sin(\tilde{a}k_{F}x)\Big[{\rm Si}\left[(1+\tilde{a})2k_{F}x\right]-{\rm Si}
\left [ (\tilde { a }
-1)2k_ {F}x\right]\nonumber \\
&&+2{\rm Si}(2k_{F}x)\Big]+{\ln}\left|\frac{1-\tilde{a}} {1+\tilde{a}} 
\right|\cos(\tilde{a}k_{F}x)  \Big\},\\
 I^{r}_{\Is}&=&2\pi I_{0}\Big[{\rm Si}((1+\tilde{a})2k_{F}x) 
-{\rm Si}\left[(\tilde{a}-1)2k_{F}x\right]\Big]  -I^{r}_{\RKKY}.
\end{eqnarray}
%\end{strip}
% 
To obtain the final results we still need to compute the contribution from the 
singularities of the integrals~(\ref{Ir}).

\subsection{Contribution from the singularities}
\noindent
To compute the contributions from the singularities we use the same method  
applied to the traditional RKKY problem in 1D~\cite{Yafet,Zawadzki}. Let us start with 
the integral 
\begin{eqnarray}
 I_{++}=I_0 \int_{-q_{+}}^{q_{+}} dq\int_{|q^\prime|>q_{+}}dq^\prime 
\frac{e^{i(q-q^\prime)k_Fx} }{q^2-{q^\prime}^2+\tilde a 
(q-q^\prime)}.\nonumber \\
\end{eqnarray}
The singularities of this integral occur when $(q-q')=0$ and $(q+q+\tilde 
a)=0$ or  $q^\prime=-\tilde{a}/2$ and $q=-\tilde{a}/2$. In the following we 
calculate the integral above around $q=q^\prime=\tilde a/2$. At this point we 
have $e^{i(q-q')k_Fx} = e^{i(-\tilde{a}+\tilde{a})k_Fx/2}=1$. Therefore,
\begin{equation}
I_{++}^{\epsilon}= \int_{-\frac{\tilde{a}}{2}-\epsilon}^{-\frac{\tilde{a}}{2} 
+\epsilon} dq\int_{-\frac{\tilde{a}}{2}-\epsilon}^{-\frac{\tilde{a}}{2} 
+\epsilon} \frac{dq'}{(q-q')(q+q'+\tilde a)},
\end{equation}
with $\epsilon\rightarrow 0^{+}$.
The integral over $q^\prime$ variable can be calculated analytically using 
\begin{eqnarray}
 \int \frac{dx}{(y-x)(y+x+a)}&=&\frac{\ln(-a-x-y)-\ln(y-x)}{2y+ a}+ \mbox{constant}.
\end{eqnarray}
Imposing the limits, after some algebraic manipulation we obtain 
\begin{eqnarray}\label{intea}
 I_{++}^{\epsilon}\!\!=\!\!\int_{-\frac{\tilde{a}}{2}-\epsilon}^{-\frac{\tilde{a}}{2} +\epsilon}\!\! dq\frac{\ln|(q+\tilde{a}/2)+\epsilon|-\ln|(q+\tilde{a}/2) 
-\epsilon|} {q+\tilde{a}/2}.
\end{eqnarray}
Performing the variable change $x=q+\tilde{a}/2$ the above integral
becomes 
\begin{equation}
 I_{++}^{\epsilon}=\int_{-\epsilon}^{\epsilon}dx\left[\frac{\ln|x+\epsilon|}{x}-
 \frac{\ln|x-\epsilon|}{x}\right].
\end{equation}
This expression can be written as 
\begin{eqnarray}\label{dig}
 I_{++}^{\epsilon}&=&\int_{-1}^{1}du\left[\frac{\ln|1+u|}{u}-
\frac{\ln|1-u|}{x}\right]=2\rm{Li}_{2}(1)-2\rm{Li}_{2}(-1)= 
\frac{\pi^{2}}{2},
\end{eqnarray}
where 
\begin{eqnarray}
\rm{Li}_{2}(x)=-\int_{0}^{x}\frac{\ln|1-t|}{t} dt
\end{eqnarray}
is the Dilogarithmic function. In the last line passage in (\ref{dig}) we 
have used\cite{Grobner} $\rm{Li}_{2}(1)=\pi^{2}/6$  and 
$\rm{Li}_{2}(-1)=-\pi^{2}/12$. 

Likewise, we can show that 
% 1
\begin{equation}\label{dig2}
I^{\epsilon}_{--}=\frac{\pi^{2}}{2}.
\end{equation}
The others two integrals render slightly different results. Let us look at 
the correction for 
\begin{equation}
 I_{+-}=I_0 \int_{-q_{+}}^{q_{+}} dq\int_{-\infty}^{\infty}dq^\prime 
\frac{e^{i(q-q^\prime)k_Fx} }{q^2-{q^\prime}^2+\tilde a 
(q+q^\prime)}.
\end{equation}
Here the contribution are accounted when $(q+q')=0$ and $(q-q'+\tilde{a})=0$, 
from which we extract $q'=\tilde{a}/2$ and $q=-\tilde{a}/2$. At this point, 
$e^{i(q-q')k_Fx} = e^{i(-\tilde{a}-\tilde{a})k_Fx/2}=e^{-i\tilde{a}k_Fx}$, so 
that
\begin{eqnarray}
 I^{\epsilon}_{+-}=e^{-i\tilde{a}k_Fx}\int_{-\frac{\tilde{a}}{2}-\epsilon}^{- 
\frac{\tilde{a}}{2}+\epsilon} dq\int_{\frac{\tilde{a}}{2}- 
\epsilon}^{\frac{\tilde{a}}{2}+\epsilon} \frac{dq^\prime}{(q+q^\prime) 
(q-q^\prime+\tilde a)}.\nonumber\\
\end{eqnarray}
Using the indefinite integral 
\begin{eqnarray}
  \int \!\!\! \frac{dx}{(y+x)(y-x+a)}\!&=&\!\frac{\ln(y+x)-\ln(-a+x-y)}{2y+ a}+ \mbox{constant},
\end{eqnarray}
we obtain
\begin{eqnarray}
I^{\epsilon}_{+-}&=&e^{-i\tilde{a}k_Fx}\nonumber\\
&&\!\!\!\!\!\!\times\int_{-\frac{\tilde{a}}{2}-\epsilon}^{
-\frac{\tilde{a}}{2}+\epsilon}\!\!\! \!\! dq\frac{\ln|(q+\tilde{a}/2)+\epsilon|\!-\!{\ln}|(q+\tilde{a}/2)-\epsilon|} {q+\tilde{a}/2}.\nonumber \\
\end{eqnarray}
Apart from the prefactor $e^{-i\tilde{a}k_Fx}$,  this is the same as in 
\ref{intea}, therefore,
\begin{equation}\label{dig3}
 I^{\epsilon}_{+-}=\frac{\pi^{2}}{2}e^{-i\tilde{a}k_Fx}.
\end{equation}
The last correction, for $I_{-+}^r$, can be obtained using same argument of 
changing $\tilde{a}\rightarrow -\tilde{a}$ in (\ref{dig3}), leading to 
\begin{equation}\label{dig4}
I^{\epsilon}_{-+}=\frac{\pi^{2}}{2}e^{i\tilde{a}k_Fx}.
\end{equation}
Collecting the results of (\ref{dig}), (\ref{dig2}), (\ref{dig3}) and 
(\ref{dig4}) we obtain the corrections for the couplings
% 
%\begin{subequations}
\begin{eqnarray}\label{corrections}
I^\epsilon_{\rm{RKKY}}&=&2\Re (I_{-+}^\epsilon+I_{+-}^{\epsilon *})= 4\pi 
I_{0}\left[\frac{\pi}{2} \cos(\tilde{a}k_{F}x)\right],\nonumber \\
\end{eqnarray}
\begin{eqnarray}
I^\epsilon_{\rm{DM}}&=&-2\Im (I_{-+}^\epsilon+I_{+-}^{\epsilon *})=- 4\pi 
I_{0}\left[\frac{\pi}{2} \sin(\tilde{a}k_{F}x)\right],\nonumber\\
\end{eqnarray}
\begin{eqnarray}
I^{\epsilon}_{\rm{Ising}}&=& 2\Re (I_{++}^\epsilon+I_{-- }^\epsilon) 
-I^\epsilon_{\rm{RKKY}}\nonumber\\
&=&4\pi I_{0}\left(\frac{\pi}{2}\right) 
-I^\epsilon_{\rm{RKKY}}.
\end{eqnarray}
%\end{subequations}
% 

We  now subtract the results of the Eqs.~(\ref{corrections}) from those of 
Eqs.~(\ref{Coupling_full}) to obtain our final analytical 
results for the indirect coupling
\begin{eqnarray}\label{Coupling_full_final}
I_{\RKKY}&=&\pi I_{0}\Big\{\sin(\tilde{a}k_{F}x)\Big[{\rm Ci}\left[
(1-\tilde{a})2k_{F} x\right]- {\rm Ci}\left[(1+\tilde{a})2k_{F}x\right]\Big] \nonumber \\
&&+\cos(\tilde{a}k_{F}x)\Big[{\rm Si}\left[(1+\tilde{a})2k_{F}x\right]-{\rm Si}
\left[(\tilde { a } -1)2k_{F}x\right]\nonumber \\
&&+2{\rm Si}(2k_{F}x)\Big]-{\ln}\left|\frac{1-\tilde{a}}{1+\tilde{a}} \right|\sin(\tilde{a}k_{F}x) 
 \Big\}-4\pi I_{0}\left[\frac{\pi}{2}\cos(\tilde{a}k_{F}x)\right],\\
 I_{\DM}&=&-\pi I_{0}\Big\{\cos(\tilde{a}k_{F}x)\Big[{\rm Ci}\left[(1+ 
\tilde{a})2k_{F}x\right]- {\rm Ci}\left[(1-\tilde{a})2k_{F}x\right]\Big]\nonumber \\ 
&&+\sin(\tilde{a}k_{F}x)\Big[{\rm Si}\left[(1+\tilde{a})2k_{F}x\right]-{\rm Si}
\left[(\tilde { a } -1)2k_ {F}x\right]\nonumber \\
&&+2{\rm Si}(2k_{F}x)\Big] +{\ln}\left|\frac{1-\tilde{a}} {1+\tilde{a}} \right|\cos(\tilde{a}k_{F}x)  
\Big\}+4\pi I_{0}\left[\frac{\pi}{2} \sin(\tilde{a}k_{F}x)\right],\\
I_{\Is}&=&2\pi I_{0}\Big[{\rm Si}\left[(1+\tilde{a})2k_{F}x\right] 
-{\rm Si}\left[(\tilde{a}-1)2k_{F}x\right]\Big] -4\pi 
I_{0}\left(\frac{\pi}{2}\right) 
-I_{\RKKY}.
\end{eqnarray}
%\end{strip}
Notice that if we take $\tilde{a}=0$ the usual result $I_{\rm{RKKY}}=4\pi 
I_{0}\left[{\rm Si}(2k_{F}x)-{\pi}/{2}\right]$ and 
$I_{\rm{DM}}=I_{\rm{Ising}}=0$ is recovered, as expected. Interestingly, 
however,  the asymptotic behavior of these expressions are
 \begin{eqnarray}
I_{\RKKY}&=&-\pi 
I_{0}{\ln}\left|\frac{k_{\rm F}-2k_{\rm R}}{k_{\rm F}+ 2k_{\rm R}}
\right|\sin(2k_{\rm R}x),\\
I_{\DM}&=&-\pi 
I_{0}{\ln}\left|\frac{k_{\rm F}-2k_{\rm R}}{k_{\rm F}+ 2k_{\rm R}} 
\right|\cos(2k_{\rm R}x),\\
 I_{\Is}&=& \,\,\,\  \!\pi 
I_{0}{\ln}\left|\frac{k_{\rm F}-2k_{\rm R}}{k_{\rm F}+ 2k_{\rm R}}
\right|\sin(2k_{\rm R}x).
\end{eqnarray}
Where we use $\lim_{x\rightarrow \infty} {\rm Ci}(x)=0$, and 
$\lim_{x\rightarrow \infty}{\rm Si}(x)=\pi/2$. These unsuppressed  
oscillations appearing in these asymptotic expressions is the principal result 
of our work.
\section{Analytical vs. numerical results}

Despite the complexities involved in obtaining the analytical results, 
numerically it is rather straightforward. Basically, we need to calculate the 
integrals (\ref{coupling_integral}) numerically. In fact, here we simply 
perform these integrals using a numerical subroutine built in {\tt Julia} 
programming language\cite{Julia}. To get convergence, as usual we add an 
infinitesimal imaginary to the denominator of (\ref{coupling_integral}) so that 
the integrals we indeed solve numerically are
\begin{figure}[h!]
\centering
\subfigure{\includegraphics[clip,width=4.5in]{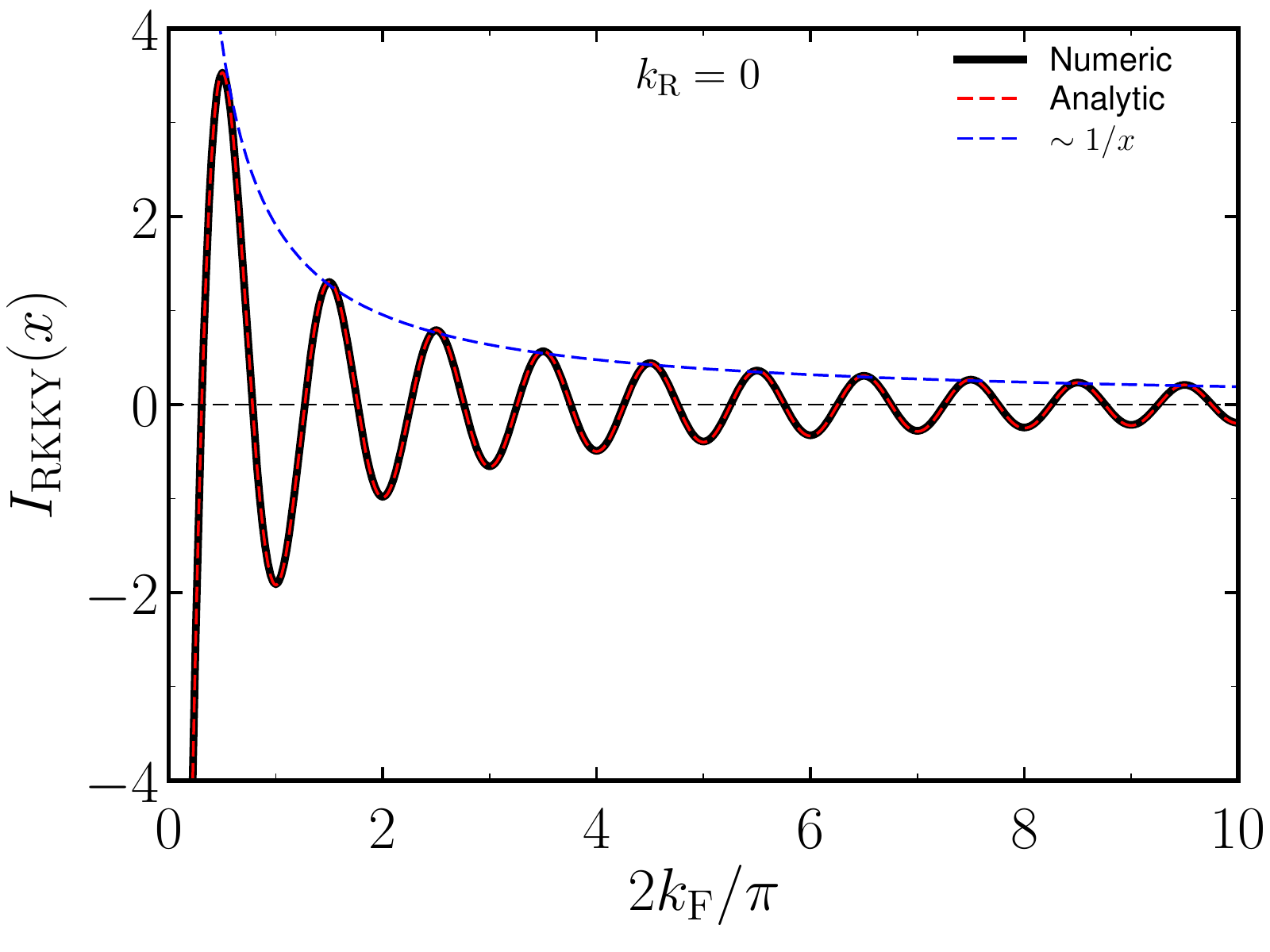}}
\caption{Comparison between the analytical and the numerical results for the  
RKKY vs  $x$ in the absence of SOC  ($a=0$). Solid black lines correspond to the numerical results while  dashed red lines correspond to the analytical results. The blue dashed line shows a function  $\sim 1/x$  to show that in the absence of the SOC the RKKY coupling indeed decays as expected.} 
\label{fig_supp1}
\end{figure} 
\begin{figure}[h]
\centering
\subfigure{\includegraphics[clip,width=4.5in]{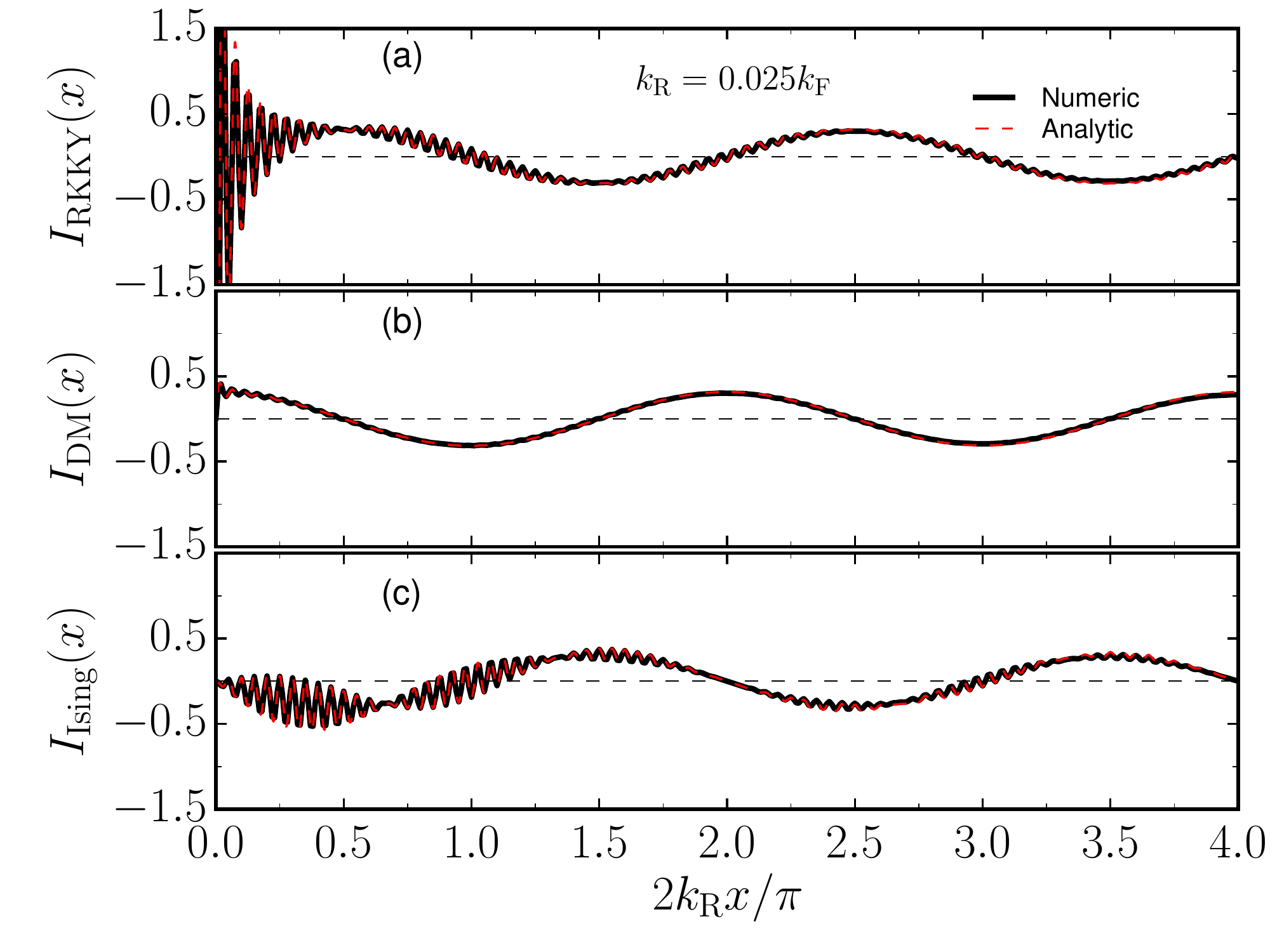}}
\caption{Comparison between the analytical and the numerical results for the 
indirect couplings.  RKKY (a)  Dzaloshinkii-Moryia (b) and Ising (c) 
couplings as a function of  $x$ for $k_{R}=0.025k_{\rm F}$. 
Solid black lines correspond to the numerical results, dashed red lines 
correspond to the analytical results and dash-dot blue lines show the 
asymptotic behavior of the couplings.} 
\label{fig_supp2}
\end{figure} 
\begin{eqnarray}\label{coupling_integral_num}
I_{\delta\nu}={J^2}\int_{-k_\delta}^{k_\delta}\frac{dk}{2\pi}\int_{ |k^\prime|>k_\delta } 
\frac{dk^\prime}{2\pi} \frac{e^{i(k-k^\prime)x}} {\frac{\hbar^{2}}{2m^*}(k^{2} - 
k^{^\prime 2})+\delta\alpha_{\rm R}(k-\delta\nu k^\prime)+i\eta},
\end{eqnarray}
with $\eta=0^+$.
The expression above is exactly the same we  obtain when we used scattering 
theory to obtain the indirect interaction via the Lippmann-Schwinger 
equation~\cite{Sakurai}, having in mind that we need to account for the Fermi 
sea and the Pauli's exclusion principle. Having calculated the integrals 
numerically, we obtain  the indirect coupling using the expressions just using 
the expressions for  $I_{\RKKY}$, $I_{\DM}$ and $I_\Is$  obtained in the end of 
Sec.~(\ref{App:Perturbation}). The analytical (dashed red line) and the numerical (solid black line) results are compared in 
Fig.~(\ref{fig_supp1}) in the absence of SOC ($a=0$) and in Fig~(\ref{fig_supp2}) 
for $k_R=0.025k_{\rm F}$. Notice that, as expected, the oscillations  
are suppressed as $1/x$, as shown by the dashed blue line.

\section{Effective Hamiltonian in terms of Green's function }\label{App:Greens_function}
In this section we present a derivation of an expression for the effective 
Inter-impurity Hamiltonian in terms of Green's function  in the position 
space for the 1D system in the presence of the spin-orbit interaction.

Let us start with Eq.~(\ref{mixed}) that can be written as
\begin{eqnarray}\label{mixed1}
\!H_{\rm eff}\!=\!\!\sum_{k,h}^{\rm occ.}\, \sum_{{k^\prime, 
h^\prime}}^{\rm empty}\left[\frac{ \langle k,h|H_1|k^\prime, 
h^\prime\rangle\langle k',h'|H_2|k,h\rangle}{\varepsilon_{kh}-
 \varepsilon_{k'h'}}\!+\!{\rm H.c}\right].\nonumber \\
 \end{eqnarray} 
Defining the retarded Green's function
\begin{eqnarray}
\hat G_{k^\prime,h^ 
\prime}(\varepsilon_{kh})=\frac{|k^\prime, 
h^\prime\rangle\langle k',h'|}{\varepsilon_{kh}- 
\varepsilon_{k^\prime h^\prime}},
\end{eqnarray}
in which $\varepsilon_{kh} \rightarrow \varepsilon_{kh} + 0^+ $, the Eq.~(\ref{mixed1}) can be 
written as
\begin{eqnarray}\label{mixed2}
H_{\rm eff}=\sum_{k,h}^{\rm occ.}\, \sum_{{k^\prime, 
h^\prime}}^{\rm empty}{ \langle k,h|H_1 \hat G_{k^\prime,h^ 
\prime}(\varepsilon_{kh})H_2|k,h\rangle}+{\rm H.c.}
 \end{eqnarray} 
Let us now introduce two closure relations in the position space, 
$\sum_{x,\sigma} |x,\sigma \rangle\langle x,\sigma| $, to obtain
%
%\begin{strip}
\begin{eqnarray}\label{H_eff_10}
\!\!\!\!\!\!\!\!\!\!\!\!\! \!\!\!\! H_{\rm eff}&=&\sum_{k,h}^{\rm occ.} \sum_{k^\prime, 
h^\prime}^{\rm empty}\sum_{xx^\prime}\sum_{\sigma \sigma^\prime} \langle 
k,h|H_1 |x,\sigma \rangle\langle x,\sigma| \hat 
G_{k^\prime,h^\prime}(\varepsilon_{kh})|x^\prime,\sigma^\prime \rangle \langle x^\prime,\sigma^\prime|H_2|k,h\rangle+{\rm H.c.}.\quad
\end{eqnarray}
%\end{strip}
% 
% 
In the real position and spin space, the Hamiltonian $H_i$ of 
Eq.~(\ref{impurit}) acquires the form 
\begin{eqnarray}
H_i&=&J \left[S^z_{x}(c^\+_{x \up} c_{x \up}-c^\+_{x \dn} 
c_{x \dn}) + S^+_{x}c^\+_{x \dn} c_{x \up}\right.\nonumber \\
&&\left.+ S^-_{x}c^\+_{x \up} c_{x \dn} \right] \delta(x-x_i)
=H(x)\delta(x-x_i).
\end{eqnarray}
Inserting this into Eq.~(\ref{H_eff_10}), after some straightforward algebraic manipulations we obtain
\begin{eqnarray}\label{H_eff_final}
H_{\rm eff} &=&J^2\sum_{h h^\prime} \langle h|\left({\bf S}_1 \cdot 
{\bf s} \right) |h^\prime\rangle 
\langle h^\prime | \left({\bf S}_2 \cdot 
{\bf s}\right)|h\rangle\nonumber \\
&&\times\sum_{k}^{\rm occ.}\sum_{k^ \prime}^{\rm empty} 
G_{k^\prime,h^\prime}(x_1,x_2,\varepsilon_{kh}) \langle 
k|x_1\rangle \langle x_2|k\rangle  +{\rm 
H.c.}.
 \end{eqnarray} 
Here we have  defined the scalar retarded Green's function
\begin{eqnarray}
G_{k^\prime,h^\prime}(x_1,x_2,\varepsilon_{kh})=\frac{\langle x_1|k^\prime \rangle \langle k^\prime |x_2\rangle}{\varepsilon_{kh}- 
\varepsilon_{k^\prime h^\prime}}.
\end{eqnarray}
We now use $\langle x_{i}|k\rangle=\frac{1}{\sqrt{N}}e^{ikx_{i}}$  and 
transform the summation into integrals we obtain
\begin{eqnarray}\label{H_eff_final}
H_{\rm eff}&=&NJ^2\sum_{h h^\prime} \langle h|\left({\bf S}_1 \cdot 
{\bf s} \right) |h^\prime\rangle \langle h^\prime | \left({\bf S}_2 \cdot 
{\bf s}\right)|h\rangle \nonumber \\
&&\!\!\!\times\int_{\rm occ.} \frac{dk}{2\pi} e^{ik(x_2-x_1)} 
\int_{\rm empty} \frac{dk^\prime}{2\pi} 
G_{k^\prime,h^\prime}(x_1,x_2,\varepsilon_{kh})+{\rm H.c.}.
 \end{eqnarray} 
As discussed in detail by Valizadeh\cite{Valizadeh}, further simplification of 
Eq.~(\ref{H_eff_final}) towards the a similar expression as  Eq.~(5) of 
Ref.~\cite{Imamura} requires changing the  order of the integrals over 
$k$ and $k^\prime$, which  may lead to spurious result in 1D case. Moreover, 
the integral over $k^\prime$ cannot be extended from $-\infty$ to $\infty$, 
since the extra contribution to the double integral does not vanish in the 
presence of spin-orbit coupling.


\begin{thebibliography}{<num>}
\bibitem{footenote1} Considering spin-$1/2$ impurities is not really necessary 
here, however, it makes our calculation simpler, as we can write 
straightforwardly write the spin operator of the impurity in terms of the 
Pauli matrices.
\bibitem{footenote2} As usual, the sine and cossine integral functions are 
respectively defined as ${\rm Ci}(x)=\int_0^x(1/t){\cos( t)} dt$ and ${\rm 
{\rm Si}}(x)=\int_0^x(1/t){\sin( t)} dt$.
\end{thebibliography}
\end{document}